# Correlation-enhanced electron-phonon coupling and superconductivity in (Ba,K)$SbO_3$ superconductors

Zhihong Yuan, Pengyu Zheng, Yiran Peng, Rui Liu, Xiaobo Ma,
Guangwei Wang, Tianye Yu, and Zhiping Yin*

*Department of Physics and Center for Advanced Quantum Studies, Beijing Normal University, Beijing 100875, China.*

The electronic structure, lattice dynamics, and electron-phonon coupling (EPC) of the newly discovered (Ba,K)$SbO_3$ superconductors are investigated by first-principles calculations. The EPC of (Ba,K)$SbO_3$ is significantly enhanced by considering non-local electronic correlation using the Heyd-Scuseria-Ernzerhof hybrid exchange-correlation functional (HSE06). The EPC strength $\lambda$ of $Ba_{0.35}K_{0.65}SbO_3$ is strongly increased from 0.33 in local-density approximation calculations to 0.59 in HSE06 calculations, resulting in a superconducting transition temperature $T_c$ of about 14.9 K, which is in excellent agreement with experimental value of ~ 15 K. Our findings suggest (Ba,K)$SbO_3$ are extraordinary conventional superconductors, where non-local electronic correlation expands the bandwidth, enhances the EPC, and boosts the $T_c$. Moreover, we find both $\lambda$ and $T_c$ depend crucially on the K-doping level for (Ba,K)$SbO_3$ and (Ba,K)$BiO_3$ compounds. (Ba,K)$SbO_3$ have stronger EPC strength and higher $T_c$ than those of (Ba,K)$BiO_3$ at the same K-doping level.

## I. INTRODUCTION

Since the discovery of superconductivity in perovskite-type bismuthates with $T_c$ up to 30 K [1,2], there have been considerable experimental [3-8] and theoretical [9-13] efforts paid to explore its superconducting mechanism.

Historically, tunneling measurements and specific heat experiments have consistently confirmed that the electron-phonon interaction plays an important role in bismuthates' superconductivity [3-5]. However, electron-phonon coupling (EPC) calculated by density functional theory (DFT) and density functional perturbation theory (DFPT) based on the standard local-density approximation (LDA) or generalized-gradient approximation (GGA) method fail to account for the unexpectedly high-$T_c$ of $(Ba,K)BiO_3$ (BKBO) [9,12,13]. In 2013, Yin *et al*. [10] found that EPC of BKBO was strongly enhanced by considering non-local electronic correlation. In their calculations, using DFT Heyd-Scuseria-Ernzerhof hybrid functional (HSE06) or self-consistent quasiparticle GW (scQPGW) [14], EPC strength of BKBO was strongly increased from the DFT-LDA value $\lambda \simeq 0.3$ to $\lambda \simeq 1$, which was strong enough to account for the high $T_c$ in BKBO. This finding was confirmed independently in 2019 by Louis group [11] using first-principles linear-response GW method. In their study, Li *et al*. [11] found that the GW self-energy renormalizes the DFT-LDA electron-phonon matrix elements and enhances the $\lambda$ from 0.47 to 1.14 for $Ba_{0.6}K_{0.4}BiO_3$. In 2018, Wen *et al*. [7] provided the first direct experimental proof that including long-range Coulomb interaction can obtain the correct band structure and EPC strength for BKBO compounds. In their experiment, they proved that $Ba_{0.51}K_{0.49}BiO_3$ is an extraordinary Bardeen-Cooper-Schrieffer (BCS) superconductor, of which $\lambda$ and $T_c$ are about $1.3 \pm 0.1$ and 22 K, respectively. Thus, the long-standing problem of superconducting mechanism in bismuthates compounds was solved.

Despite the mystery of superconducting mechanism of BKBO compounds has been solved by considering long-range Coulomb interactions, using this mechanism to reliably evaluate the EPC strength of more materials is of fundamental importance for both understanding the underlying physics and designing novel functional materials. Very recently, the superconducting antimonate (Ba,K)$SbO_3$ (BKSO) was stabilized for the first time by Kim *et al*. [15] using a high pressure and high temperature synthesis route. Their studies showed that the parent compound $BaSbO_3$ is similar to $BaBiO_3$, which is also a charge density wave (CDW) insulator. The substitution of Ba by K suppresses the CDW order and leads to a cubic perovskite structure when K-doping level reaches 65 %. Moreover, superconductivity emerges at 65 % K doping level with a $T_c$ of about 15 K, which is lower than the maximum $T_c$ of about 30 K of 40% K-doped BKBO. The EPC strength of BKSO and whether BKSO has a conventional superconducting mechanism or a superconducting mechanism similar to BKBO remain to be explored.

In this work, our main focus is to explore the superconducting mechanism of this newly discovered superconductor family BKSO. Given that Bi and Sb have similar chemical properties, and both superconducting BKSO and BKBO have the ideal cubic perovskite structures, we believe that considering non-local electronic correlation in calculations could describe the properties of the metallic BKSO compounds more accurately. Therefore, the EPC in BKSO compounds is evaluated by combining DFPT-LDA calculations and DFT-HSE06 supercell calculations. The second object of this work is to do a detailed investigation on the effects of K-doping level on the EPC and $T_c$ of BKSO compounds, and to compare the related properties with BKBO compounds.

The rest of the paper is organized as following: In Section II, the computational methods and details used in this work are described. Section III A reports the crystal structure, electronic structure, lattice dynamics and the EPC of $Ba_{0.35}K_{0.65}SbO_3$. Non-local electronic correlation is included to estimate the realistic $\lambda$ and $T_c$. Section III B discusses the K-doping dependence of crystal structure, electronic structure, EPC and $T_c$ of BKSO. Section III C presents the K-doping dependence of EPC and $T_c$ of BKBO for comparison with the results of BKSO. In Section IV, a summary of this work is given.

## II. METHODS

### A. Computational theory

Based on the EPC theory [16,17], the EPC strength $\lambda$ can be determined by the summation of the mode- and momentum-dependent coupling constant $\lambda_{qv}$

$$\lambda = \frac{1}{N_q}\sum_{qv}\lambda_{qv}, \tag{1}$$

where $N_q$ is the number of $q$ points, the $\lambda_{qv}$ associated with a specific phonon mode and wave vector $q$ is given by

$$\lambda_{qv} = \frac{2}{N(\varepsilon_F)}\sum_k \frac{1}{\omega_{qv}}\left|M^{\nu}_{k,k+q}\right|^2 \delta(\varepsilon_k - \varepsilon_F)\delta\left(\varepsilon_{k+q} - \varepsilon_F\right), \tag{2}$$

here, $N(\varepsilon_F)$ is the electronic density of states (DOS) per spin at the Fermi level $\varepsilon_F$, $\omega_{qv}$ represents the phonon frequency of branch $v$ with wave vector $q$, and $M^{\nu}_{k,k+q}$ are the electron-phonon matrix elements given by the following formula:

$$M^{\nu}_{k,k+q} = \sum_j \left(\frac{\hbar^2}{2M_j\omega_{qv}}\right)^{1/2} \epsilon^{\nu}_{q,j} \cdot \left\langle k+q \left| \frac{\delta V}{\delta u^{\nu}_{q,j}} \right| k \right\rangle, \tag{3}$$

where $j$ runs over the atoms in the unit cell and $\frac{\delta V}{\delta u^{\nu}_{q,j}}$ is the partial derivative of the total Kohn-Sham potential energy relative to a given phonon displacement $u^{\nu}_{q,j}$ of the $j$th atom.

Following the discussion in Ref. [10], for optical vibration modes, the electron-phonon matrix element of a wave vector $q$ commensurate with the lattice can be deduced from the shifts of the energy bands in a supercell calculation, which was denoted as the reduced electron-phonon matrix element (REPME) $D^{\nu}_{k,q}$. For states on the Fermi surface, the electron-phonon matrix elements can be read directly from the splitting of the energy bands caused by the atomic displacements of the corresponding phonon mode. The realistic EPC can be estimated from a DFPT-LDA or DFPT-GGA calculation by rescaling the LDA or GGA REPMEs to the actual values given by more advanced approaches while keeping the integral over $k$ and $q$ of Eq. (2) at the LDA or GGA level (Ignoring the small modifications of the $N(\varepsilon_F)$ and the $\omega_{q\nu}$ in HSE06 or GW calculations). In practice, a good estimation is achieved by using both an advanced approach and LDA or GGA to evaluate the REPMEs for all the strongly coupled phonon modes (which can be obtained from the DFPT-LDA or DFPT-GGA calculations) at special points in the Brillouin zone,

$$\lambda_H = \sum_{\nu} \lambda_{H\nu} \simeq \sum_{\nu} \lambda_{L\nu} \langle |D^{\nu}_{H}|^2 / |D^{\nu}_{L}|^2 \rangle, \quad (4)$$

where HSE06 method is denoted as $H$ and the LDA or GGA is denoted as $L$. If the enhancements in the REPMEs of all the strongly coupled branches are of comparable magnitude, the $\lambda_H$ can be estimated by

$$\lambda_H = \lambda_L \langle |D^{\nu}_{H}|^2 / |D^{\nu}_{L}|^2 \rangle, \quad (5)$$

and the logarithmic average frequency $\omega_{\log, H}$ can be estimated from the corresponding LDA value via an empirical relation as

$$\omega_{\log,H} \simeq \omega_{\log, L} \ (1+\lambda_L)^{1/2} / (1+\lambda_H)^{1/2}, \quad (6)$$

the $T_c$ is calculated by using the modified McMillan equation

$$T_c = \frac{\omega_{\log}}{1.20} exp\left(-\frac{1.04(1+\lambda)}{\lambda-\mu^*(1+0.62\lambda)}\right). \quad (7)$$

where $\mu^*$ is the effective Coulomb repulsion parameter.

## B. Computational details

In our calculations, the ideal cubic perovskite structures are adopted for all doped materials. The lattice constants are optimized by both VASP [18] and Quantum ESPRESSO (QE) [19] using the LDA functional [20] at different doping levels as shown in Table I. We use the optimized lattice constants given by VASP [18] in all calculations including DFT-LDA [20], DFT-HSE06 [14] and DFPT-LDA [21] calculations since they agree better with available experimental values [8, 15, 22]. The band structures, Fermi surfaces and electronic density of states at all doping levels presented in this work are calculated by VASP code. The Fermi surface is sampled using the Gaussian smearing method with a 0.01 eV smearing width. The electronic density of states is calculated using the tetrahedron method. For VASP calculations, the energy cutoff of 600 eV is used. The DFPT-LDA calculations at all doping levels are simulated by using the QE [19] code with Norm-conserving [23] pseudopotential. The kinetic energy cutoff and the charge density cutoff of the plane wave basis are chosen to be 100 Ry and 400 Ry, respectively. The EPC constant is obtained with $18 \times 18 \times 18$ $k$ mesh and $6 \times 6 \times 6$ $q$ mesh with 20 independent $q$ points in the irreducible Brillouin zone.

To simulate the substitution of Ba by K, the self-consistent virtual-crystal approximation (VCA) [12, 24] method is used, in which the elemental ionic pseudopotentials of Ba and K are combined to construct the virtual pseudopotential of the virtual atom $Ba_{1-x}K_x$, i.e. $V(\mathrm{Ba}_{1-x}\mathrm{K}_x) = (1-x)V(\mathrm{Ba}) + xV(\mathrm{K})$.

In order to calculate REPMEs of some important phonon modes, we calculate band structures of supercells that are adaptive to the momentum $q$ of the phonon mode [10]. In details, for oxygen-oscillating mode at the X point [Fig. 1(a)] and the oxygen-stretching mode at the M point [Fig. 1(b)], we plot the band structure in the Brillouin zone of the tetragonal unit cells corresponding to the 2 × 1 × 1 and 2 × 2 × 1 supercells of the simple cubic unit cell with its oxygen-oscillating and oxygen-stretching distortions, respectively. For oxygen-breathing mode at the R point [Fig. 1(c)], we plot the band structure in the Brillouin zone of the face-centered cubic unit cell, corresponding to a 2 × 2 × 2 supercell of the simple cubic unit cell with its oxygen-breathing distortions.

## III. RESULTS AND DISCUSSION

### A. The properties of $Ba_{0.35}K_{0.65}SbO_3$

#### 1. Crystal structure

The powder x-ray diffraction (XRD) pattern [15] showed that 0.65K doped $BaSbO_3$ ($Ba_{0.35}K_{0.65}SbO_3$) crystallizes in a simple cubic perovskite structure, as shown in Fig. 1. The calculated lattice constants of $Ba_{0.35}K_{0.65}SbO_3$ are $a = b = c = 4.046$ Å, which is in good agreement with the recent experimental value of 4.067 Å [15]. The parent compound $BaSbO_3$ is a robust insulator with a large CDW gap of 2.54 eV [15]. Rietveld refinement of neutron powder diffraction data has confirmed the parent compound $BaSbO_3$ crystallizes in cubic symmetry (space group *Fm*-3*m*) with oxygen-breathing distortion along nearest-neighbor Sb-O bonds (Fig. 1c) from the ideal perovskite structure. The substitution of Ba by K suppresses CDW and leads to the cubic perovskite structure when K-doping level reaches 65 % [15].

#### 2. Electronic structures

Figure 2(a) presents the orbital-resolved electronic band structure as well as the total and partial density of states (DOS) of $Ba_{0.35}K_{0.65}SbO_3$, using the HSE06 hybrid functional in the DFT framework. The conduction band that crosses the Fermi level gives rise to a Fermi surface in the shape of rounded cube around the Γ point (in Fig. 2(c)). Orbital analysis indicates that O-2*p* and Sb-5*s* orbitals play a dominant role around the Fermi level, which is consistent with previous calculations [15]. The partial DOS reveals that the valence bands from -3.3 to -13 eV mainly consist of O-2*p* orbitals and a small amount of Sb-5*s* and Sb-5*p* orbitals.

In order to show the differences of electronic structures between LDA functional and HSE06 hybrid functional, the band structures and DOSs using both functionals are presented in Fig. 2(b). The band structures (left of Fig. 2(b)) indicate that the bandwidth of the conduction band in HSE06 calculation is significantly wider than that in LDA calculation. For example, along the Γ-X (Γ-M) direction, the bandwidth of the conduction band is 32% (26%) larger in HSE06 calculation than in LDA calculation. From the DOS plot (right of Fig. 2(b)), there is a clear difference in the locations of the peaks below the Fermi level. For example, the first peak is centered at about -4.5 eV in HSE06 calculation, whereas in LDA calculation it is located at about -3.0eV, which is 1.5eV away from the HSE06 result. Therefore, HSE06 hybrid functional not only expands the conduction band but also shift the position of the valence bands of the $Ba_{0.35}K_{0.65}SbO_3$ compound.

## 3. Lattice dynamical and electron-phonon coupling

We now report our main results of lattice dynamical and EPC properties of $Ba_{0.35}K_{0.65}SbO_3$ calculated by DFPT-LDA calculations. The calculated phonon dispersion curves, the Eliashberg phonon spectral function $\alpha^2F(\omega)$ as well as the cumulative frequency dependence of EPC strength $\lambda(\omega)$ are plotted in Fig. 3, where the size of red dots superimposed on the phonon dispersion curves is proportional to the mode- and momentum-dependent EPC strength $\lambda_{q\nu}$. The absence of imaginary frequency modes indicates the $Ba_{0.35}K_{0.65}SbO_3$ compound in the cubic perovskite structure is dynamically stable at the DFPT-LDA level. Three high-frequency optical branches above 470 $cm^{-1}$ are well separated from the other modes. From the total and partial phonon density of states [Fig. 3(c)], the high-frequency modes are mostly related to the vibrations of O atoms, while the low-frequency phonons are mainly contributed by the vibrations of Ba/K atoms. The intermediate-frequency phonons are primary from the vibration of O and Sb atom.

In Fig. 3(a), it can be seen the total EPC are mainly contributed by the oxygen-oscillating mode around the X point [Fig. 1(a)] and oxygen-stretching mode around the M point [Fig. 1(b)], which is further supported by the Eliashberg function $\alpha^2F(\omega)$ exhibited in Fig. 3(b). In addition, Figs. 3(a) and 3(b) both show that acoustic phonon modes have negligible contribution to the total EPC. Thus, we consider it is suitable to use Eq. (5) to estimate the realistic EPC strength from DFPT-LDA calculations in $Ba_{0.35}K_{0.65}SbO_3$.

## 4. Band structures and reduced electron-phonon matrix elements

To estimate the realistic EPC strength of $Ba_{0.35}K_{0.65}SbO_3$, we compute the REPMEs of oxygen-oscillating mode at the X point [Fig. 1(a)] and oxygen-stretching mode at the M point [Fig. 1(b)], which are the two most important phonon modes suggested by DFPT-LDA calculations above. In both cases, the displacement about 0.040 Å of each oxygen atom is employed to calculate the band structures using both the LDA and HSE06 hybrid functionals in the DFT framework.

The calculated band structure of $Ba_{0.35}K_{0.65}SbO_3$ with and without the oxygen displacement is shown in Fig. 4, where the blue arrows indicate the band splitting caused by the oxygen displacement. We find that the material remains metallic with the assumed oxygen distortion. Upon an oxygen displacement of about 0.040 Å, the band splitting values of the oxygen-oscillating mode at the X point (oxygen-stretching mode at the M point) are about 0.57 (0.81) eV in the LDA functional and 0.76 (1.08) eV in the HSE06 hybrid functional, which results in REPMEs of about 7.04 (10.01) eV/Å and 9.39 (13.35) eV/Å, respectively. The EPC strength for oxygen-oscillating mode at the X point (oxygen-stretching mode at the M point) is strongly enhanced by a factor of about 1.78 (1.78) in the HSE06 treatment, in comparison to the LDA results.

### 5. Realistic electron-phonon coupling and $T_c$

To estimate the realistic EPC of $Ba_{0.35}K_{0.65}SbO_3$, the REPMEs of the two most important phonon modes discussed above are summarized in Table III. For these two phonon modes, we have $\langle |D_H^\nu|^2/|D_L^\nu|^2\rangle \backsimeq 1.78$. Based on the DFPT-LDA result of EPC strength $\lambda_L \simeq 0.33$ and the logarithmic average frequency $\omega_{\log,L} \simeq 623\,\text{K}$, the realistic EPC strength $\lambda_H$ and $\omega_{\log,H}$ of $Ba_{0.35}K_{0.65}SbO_3$ estimated by Eq. (5) and (6) are about 0.59 and 572 K, respectively. By using the modified McMillan equation (Eq. (7)) with $\mu^* =$ 0.1, the $T_c$ of $Ba_{0.35}K_{0.65}SbO_3$ is estimated to be 14.9 K, which is in good agreement with the experimental value of about 15 K. It suggests that it is crucial to include non-local electronic correlation in order to calculate accurately the EPC strength and superconducting transition temperature of BKSO, similar to the case of BKBO.

## B. The doping dependence of properties in BKSO

### 1. Electronic structures

In this section, we further study the doping dependence of the electronic structures, lattice dynamics and EPC in $Ba_{1-x}K_xSbO_3$ ($x$ = 0.5, 0.6, 0.7, 0.8, $x$ represents the K-doping level). In all doping cases, the cubic perovskite phase structure is adopted, and the VCA method is used to simulate the K-doping. The calculated lattice constant $a$ of each doping case is listed in Table I. DFT-LDA correctly predicts the decrease of lattice constant with increasing of K-doping level, in agreement with experimental observations [15].

Figure 5 displays the band structures of $Ba_{1-x}K_xSbO_3$ ($x$ = 0.5, 0.6, 0.7, 0.8), using both the LDA and the HSE06 hybrid functionals in the DFT framework. With the increase of $x$, the most obvious change is a slight shift of the Fermi level relative to the conduction band. The comparison of the band structures obtained by LDA and HSE06 hybrid functionals shows that in all doping cases, the bandwidth of the conduction band is significantly broadened by considering non-local electronic correlation. The detailed results of the conduction band bandwidths along Γ-X and Γ-M are presented in Table II. As $x$ increases from 0.5 to 0.8, the broadening factor of bandwidth along Γ-X (Γ-M) gradually decreases from 39% (31%) to 19% (18%). This trend indicates that the influence of non-local electronic correlation is weaken with the increase of K-doping level.

The electronic DOS depicted in Fig. 6 suggests that all of these doped compounds have low electronic DOS at the Fermi level ($N(\varepsilon_F)$), and the value of $N(\varepsilon_F)$ slightly decreases as $x$ increases (insets of Fig. 6). It is also noted that there is a large difference of about 1.5 eV in the peak position of the oxygen valence states between LDA and HSE06 hybrid functional results for all doping cases. In addition, $N(\varepsilon_F)$ is reduced by about 15% in HSE06 calculation compared to LDA calculation at the same K-doping level. The shape of Fermi surface is almost identical in all cases, while the size of the rounded cube centered at the Γ point shrinks with increasing $x$. The reduced $N(\varepsilon_F)$ could suppress $T_c$ with increasing K-doing level.

## 2. Lattice dynamics and electron-phonon coupling

The calculated phonon dispersion curves with EPC strength $\lambda_{qv}$ and the $\alpha^2F(\omega)$ as well as the $\lambda(\omega)$ of $Ba_{1-x}K_xSbO_3$ ($x$ = 0.5, 0.6, 0.7, 0.8) are plotted in Fig. 7. The absence of imaginary frequency modes in Fig. 7. suggests these doped compounds in the ideal cubic perovskite structure are dynamically stable at the DFPT-LDA level based on VCA method. Note that the ideal cubic perovskite structures were stabilized in the latest experiment for $x$ = 0.65-0.8 [15]. The acoustic phonon modes have very little contribution to the total EPC for all doping cases. Therefore, the Eq. (5) is used to evaluate the realistic EPC strength of $Ba_{1-x}K_xSbO_3$ ($x$ = 0.5, 0.6, 0.7, 0.8) from DFPT-LDA calculations.

As a general trend with decreasing doping level from $x$ = 0.8 to 0.5, the high-frequency optical phonon modes contributed by vibrations of O atoms show a softening behavior, leading to the phonon gap gradually disappearing at $x$ = 0.5, which can also be seen in the phonon density of states (Fig. 7(i)-(l)). The most noticeable softening in phonon frequency is found for the oxygen-stretching mode at the M point, which strongly softens from about 681 $cm^{-1}$ at $x$ = 0.8 to only about 256 $cm^{-1}$ at $x$ = 0.5. In addition, there are apparent softening of oxygen-oscillating mode at the X point and oxygen-breathing mode at the R point, from about 557 and 724 $cm^{-1}$ at $x$ = 0.8 to only about 437 and 449 $cm^{-1}$ at $x$ = 0.5, respectively. The mode- and momentum-dependent EPC strength $\lambda_{qv}$ presented in Fig. 7(a)-(d) indicate that the stronger softening of these phonon modes corresponding to the stronger EPC of these phonon modes. As a consequence, the total EPC strength $\lambda$ increase as the K doping level decreases. At even lower doping level, the EPC becomes so strong that the phonon frequencies of these oxygen vibrational modes are imaginary, which indicates the ideal cubic perovskite structure is unstable and the materials want to be in the CDW phase.

In Fig. 7(e)-(h), the $\alpha^2F(\omega)$ show strong peaks in the high frequency region, resulting in a rapid increase of $\lambda(\omega)$ in this region. With increasing doping level from $x$ =0.5 to 0.8 (Fig. 7(e) to (h)), the main peaks in $\alpha^2F(\omega)$ shift to higher frequency monotonically while the area below the peaks is reduced at the same time. It indicates that the EPC strength decreases with increasing $x$, which is consistent with the evolution of the $\lambda(\omega)$ plot. In addition, it can be seen form the phonon density of states (Figs. 7(i) to (l)) that phonons with frequency above 400 cm$^{-1}$ mainly come from the vibrations of O atoms, while phonons with frequency below 400 cm$^{-1}$ involve the vibrations of all atoms.

In order to estimate the realistic EPC strength of $Ba_{1-x}K_xSbO_3$ ($x$ = 0.5, 0.6, 0.7, 0.8), the REPMEs of their most important phonon modes are calculated. At $x$ = 0.5, we calculate the REPMEs of three important phonon modes, namely, the oxygen-oscillating mode at the X point, the oxygen-stretching mode at the M point, and the oxygen-breathing mode at the R point. For other doping cases ($x$ = 0.6, 0.7 and 0.8, see Figs. 7(b), (c) and (d), respectively), we evaluate the REPMEs of the oxygen-oscillating mode at the X point and the oxygen-stretching mode at the M point.

Figs. S1–S4 in the Supplementary Materials [25] depict the band structures of $Ba_{1-x}K_xSbO_3$ ($x$ = 0.5, 0.6, 0.7, 0.8) with and without the oxygen displacement, using both the LDA and the HSE06 hybrid functionals. In each doping case, the displacement of each oxygen atom is 1% of its lattice constant $a$. The band splittings and the corresponding REPMEs are summarized in Table III and Fig. 8. In all cases, the band splittings and the resulting REPMEs are significantly enhanced by HSE06 hybrid functional. The band splittings and the resulting REPMEs in both LDA and HSE06 hybrid functionals increase with the increase of K doping level for the oxygen-oscillating mode at the X point. For the oxygen-stretching mode at the M point, with the increase of K doping level, the band splittings and the resulting REPMEs in LDA calculations increases, whereas they exhibit a very small nonmonotonic change in HSE06 calculations.

Based on the REPME results in Table III, we obtain the enhancement factor of $\lambda$ by HSE06 for each doping level, which are listed in Table IV. The EPC strength is strongly enhanced by the HSE06, and we find the enhancement factor of $\lambda$ gradually decreases with increasing K-doping level, which follows a consistent trend of the broadening factor of the conduction band bandwidth. Combining with the DFPT-LDA results, we estimate the $\lambda_H$ and $\omega_{\log,H}$ values of $Ba_{1-x}K_xSbO_3$ ($x$ = 0.5, 0.6, 0.7, 0.8). The $T_c$ of these doped compounds are estimated by using the modified McMillan equation (Eq. (7)) with $\mu^*$= 0.1. All these results are listed in Table IV. It can be seen that with the increase of K doping level, $\omega_{\log}$ shows an increasing trend, but $\lambda$ and $T_c$ both show a decreasing trend, which is in good agreement with the experimental observation by Kim *et al*. [15]

### C. Comparison of BKSO and BKBO

In this section, the doping dependence of EPC strength and $T_c$ in $Ba_{1-x}K_xBiO_3$ ($x$ = 0.4, 0.5, 0.6) are also investigated for comparison. Our DFT-LDA calculations suggest that $Ba_{1-x}K_xBiO_3$ is in an insulating CDW phase at $x < 0.4$. DFPT-LDA calculations of three doped compounds ($x$ = 0.4, 0.5, 0.6) show that oxygen-oscillating mode around the X point and oxygen-stretching mode around the M point, as well as the oxygen-breathing mode around the R point have large contributions to the EPC, thus we evaluate the REPMEs associated with these three phonon modes. The lattice parameter *a* also show a decreasing trend with the increase of K-doping level (see Table I), and they are in agreement with experimental values [23]. Since the atomic radius of Bi is larger than that of Sb, the lattice parameter *a* of BKBO is larger than that of BKSO at the same K-doping level.

The band splittings and the corresponding REPMEs of the three important phonon modes are presented in Table V and Fig. 9. Similar to BKSO, the band splittings and the resulting REPMEs are significantly enhanced by HSE06 hybrid functional. The band splittings and the resulting REPMEs in both LDA and HSE06 hybrid functional increase with increasing K doping level for the oxygen-oscillating mode at the X point and oxygen-stretching mode at the M point. For the oxygen-breathing mode at the R point, with the increase of K doping level, the band splittings and the resulting REPMEs increase in LDA calculations, but slightly decrease in HSE06 calculations.

The $\lambda_H$ and $\omega_{\log,H}$ values of $Ba_{1-x}K_xBiO_3$ ($x$ = 0.4, 0.5, 0.6) are obtained by combining the DFPT-LDA and REMPEs results. These results are listed in Table VI together with the calculated and experimental $T_c$ of these doped compounds. In pervious theoretical works, the $\lambda$ of $Ba_{0.6}K_{0.4}BiO_3$ is about 0.34 ~ 0.47 by using the DFPT-LDA [10,12] or DFPT-GGA [11] with different parameters, while $\lambda$ is enhanced by about a factor of 3 using HSE06 and scQPGW methods [10] and 2.4 using GW perturbation theory (GWPT) [11]. The agreement between our results ($\lambda$ ~ 0.46, enhancement factor ~ 2.03) and these previous estimates [10,11,12] is reasonable considering that precision of this type of approach and some approximations and different parameters used in calculations.

For better comparison, the variations of $\lambda$ and $T_c$ vs K-doping level for BKSO and BKBO are shown in Fig. 10. For both $Ba_{1-x}K_xSbO_3$ ($x$ = 0.5, 0.6, 0.65, 0.7, 0.8) and $Ba_{1-x}K_xBiO_3$ ($x$ = 0.4, 0.5, 0.6), the EPC strength is strongly enhanced by the HSE06 hybrid functional from the value calculated by LDA functional. The enhancement factor of $\lambda$ slightly decreases with increasing K-doping level. The K-doping level of both compounds has a significant impact on $\lambda$ and $T_c$. Moreover, both compounds share a common tendency that $\lambda$ and $T_c$ gradually decrease with increasing K-doping level from 0.5 to 0.8 for BKSO and 0.4 to 0.6 for BKBO, which is related to the decrease of $N(\varepsilon_F)$. In addition, we find BKSO has stronger EPC strength and higher $T_c$ than those of BKBO at the same K-doping level.

## IV. SUMMARY

In conclusion, the electronic structure, lattice dynamics, and EPC of the newly discovered superconductors $Ba_{1-x}K_xSbO_3$ are investigated by first-principles calculations. In order to evaluate the realistic EPC, we combine DFPT-LDA calculations and supercell calculations using a more accurate method DFT-HSE06. Our results show that, comparing to LDA, considering non-local electronic correlation not only expands the bandwidth of the conduction band and shift the energy levels of the O-2$p$ orbitals of BKSO, but also has strong impacts on the lattice dynamics and electron phonon coupling. The coupling of electrons to lattice vibrations is strongly enhanced by HSE06. The enhanced EPC is strong enough to account for the high $T_c$ superconductivity in BKSO. The EPC strength $\lambda$ of $Ba_{0.35}K_{0.65}SbO_3$ is strongly enhanced from the LDA value of 0.33 to 0.59 estimated by HSE06. Using the modified McMillan formula with $\lambda \sim 0.59$, $\omega_{\log} \sim 572$ K, and $\mu^* = 0.1$, the HSE06 calculated $T_c$ is about 14.9 K, which is in excellent agreement of the experimental value of about 15 K for $Ba_{0.35}K_{0.65}SbO_3$. Therefore, the newly discovered BKSO superconductors are extraordinary Bardeen-Cooper-Schrieffer superconductors, where non-local electronic correlation plays a crucial role to generate the required strong electron-phonon coupling.

We further explore the doping dependence of EPC strength $\lambda$ and $T_c$ in BKSO and compare these results with BKBO compounds. For each doped compound, $\lambda$ is strongly enhanced by the HSE06 hybrid functional from the value calculated by LDA functional. We find a strong K-doping dependence of $\lambda$ and $T_c$ for both compounds. BKSO has stronger EPC strength and higher $T_c$ than those of BKBO at the same K-doping level. The $T_c$ of BKSO family could surpass that of BKBO if metallic BKSO can be stabilized at a lower K-doping level.

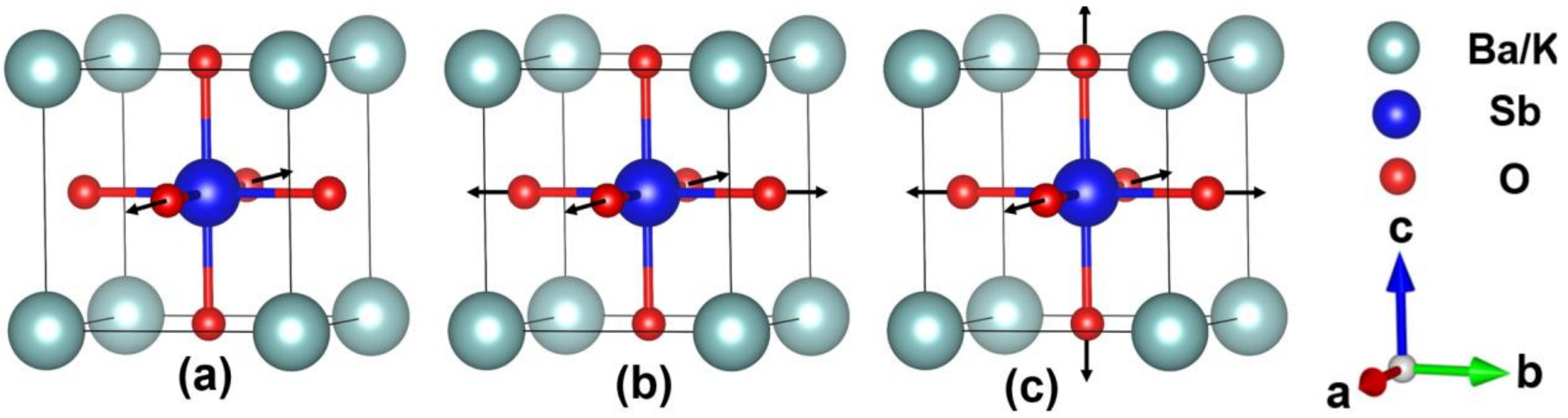

FIG. 1. Crystal structure of $Ba_{1-x}K_xSbO_3$ in the cubic perovskite phase. The arrows show (a) the oxygen-oscillating mode at the X point (b) the oxygen-stretching mode at the M point and (c) the oxygen breathing mode at the R point, respectively.

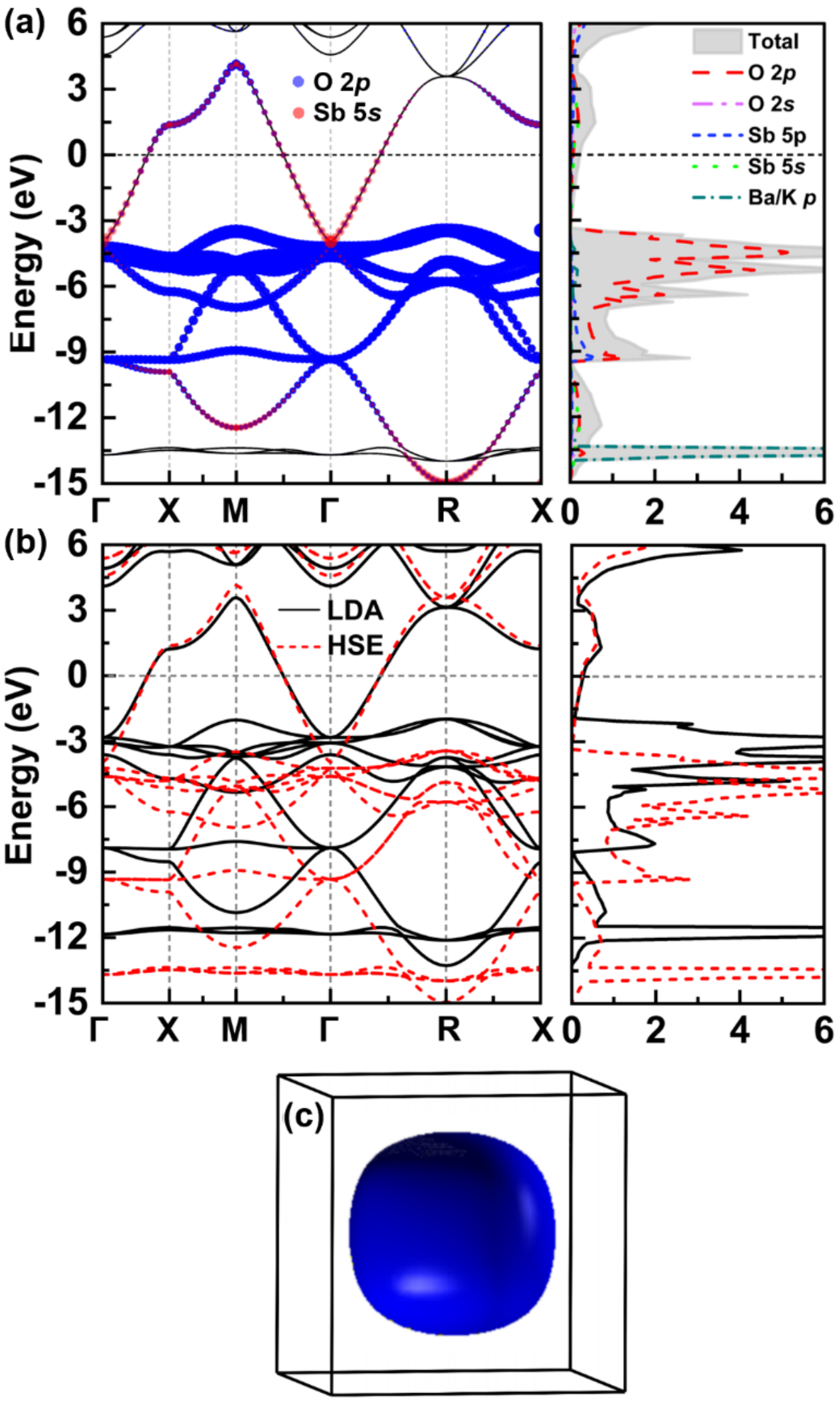

FIG. 2. (a) Band structure (left) and electronic density of states (right) of $Ba_{0.35}K_{0.65}SbO_3$ using HSE06 hybrid functional. The red and blue circles in the band structure are used to denote the Sb 5*s* and O 2*p* orbital character in different bands, where the radius of the circle is proportional to the orbital weight. The black horizontal dotted line represents the Fermi level. (b) Band structure (left) and electronic density of states (right) of $Ba_{0.35}K_{0.65}SbO_3$ using both LDA and HSE06 hybrid functionals. (c) DFT-LDA Fermi surface of $Ba_{0.35}K_{0.65}SbO_3$.

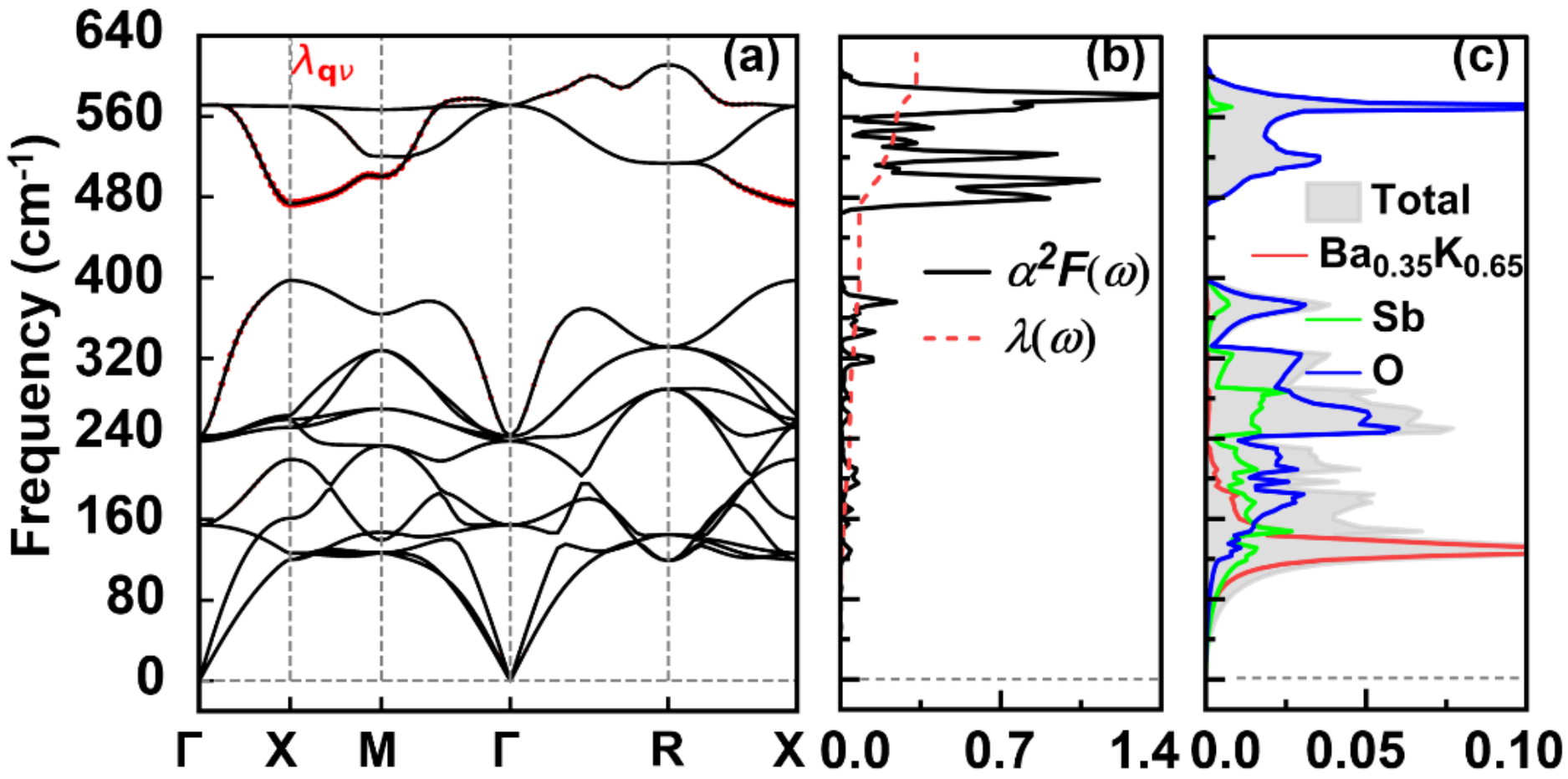

FIG. 3. The DFPT-LDA calculated lattice dynamics and EPC of $Ba_{0.35}K_{0.65}SbO_3$. (a) Phonon spectra. The radius of the red dots is proportional to the mode- and momentum-dependent electron-phonon-coupling strength $\lambda_{qv}$. (b) the Eliashberg function $\alpha^2F(\omega)$. (c) The corresponding phonon density of states.

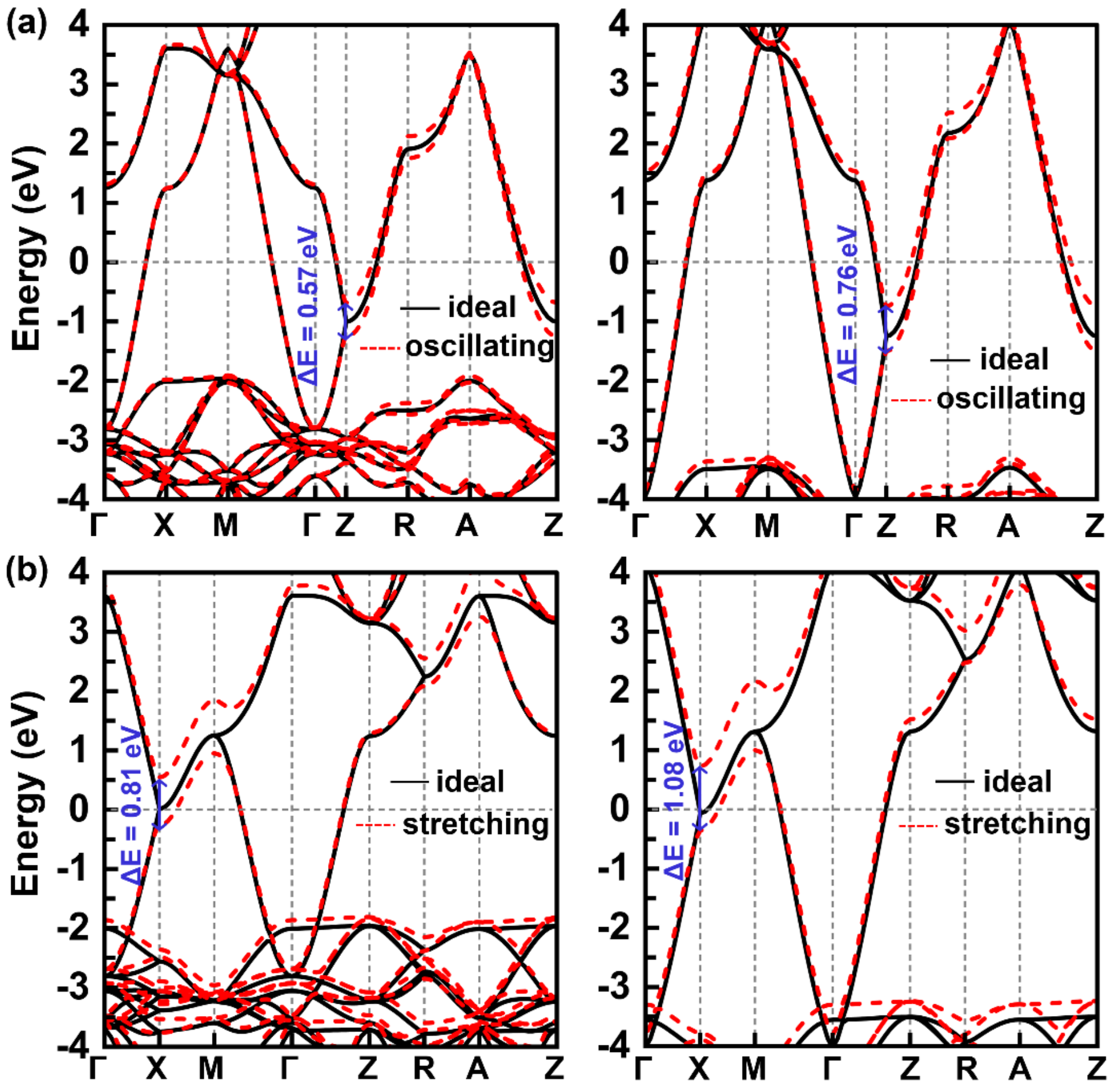

FIG. 4. Illustration of the REPMEs of (a) oxygen-oscillating mode at the X point and (b) oxygen-stretching mode at the M point in $Ba_{0.35}K_{0.65}SbO_3$. The band structures with and without the oxygen displacement are calculated by DFT using both the LDA (left) and HSE06 (right) hybrid functionals.

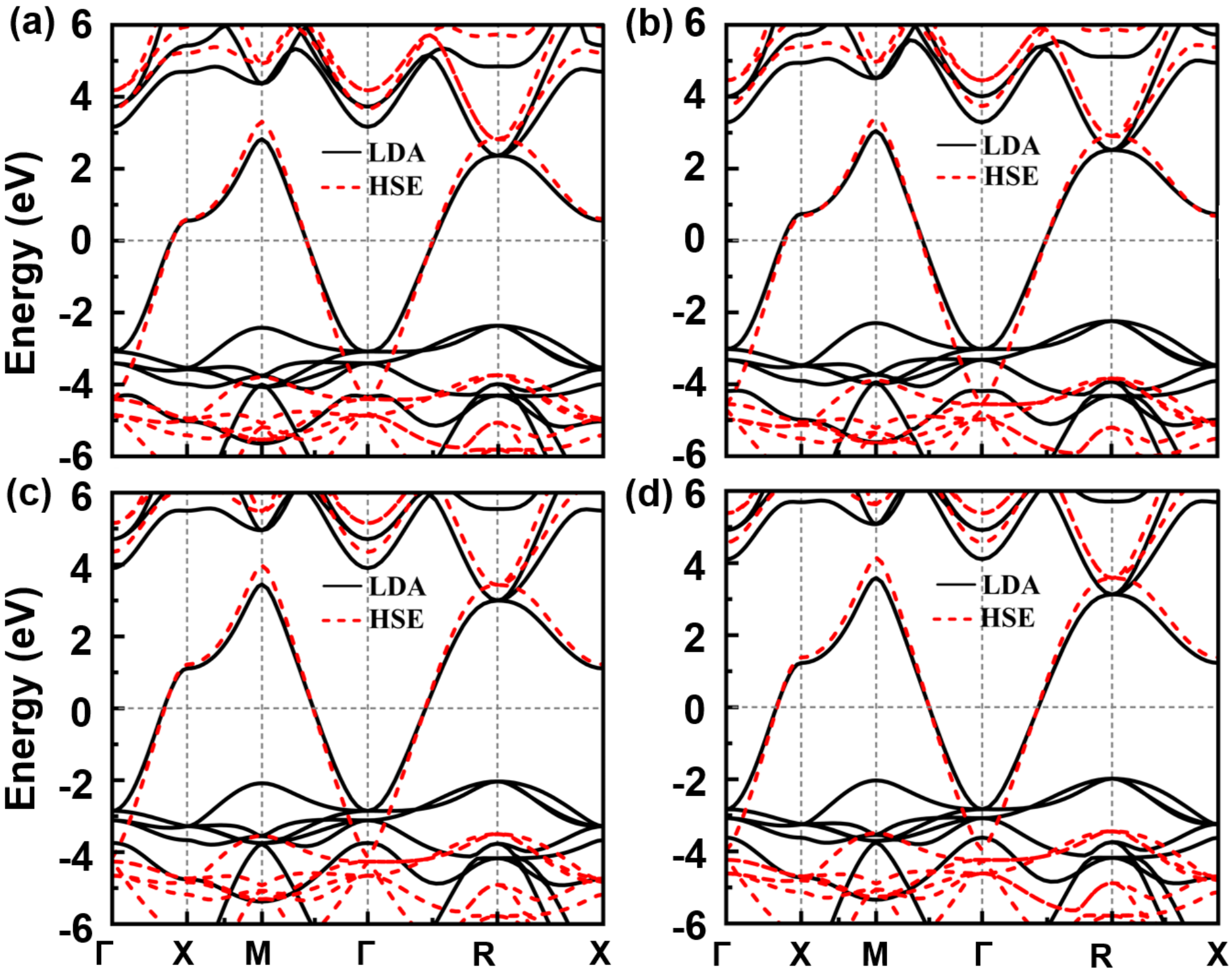

FIG. 5. Band structures of $Ba_{1-x}K_xSbO_3$ by using both the LDA and the HSE06 hybrid functionals, (a) $x$ = 0.5, (b) $x$ = 0.6, (c) $x$ = 0.7, (d) $x$ =0.8.

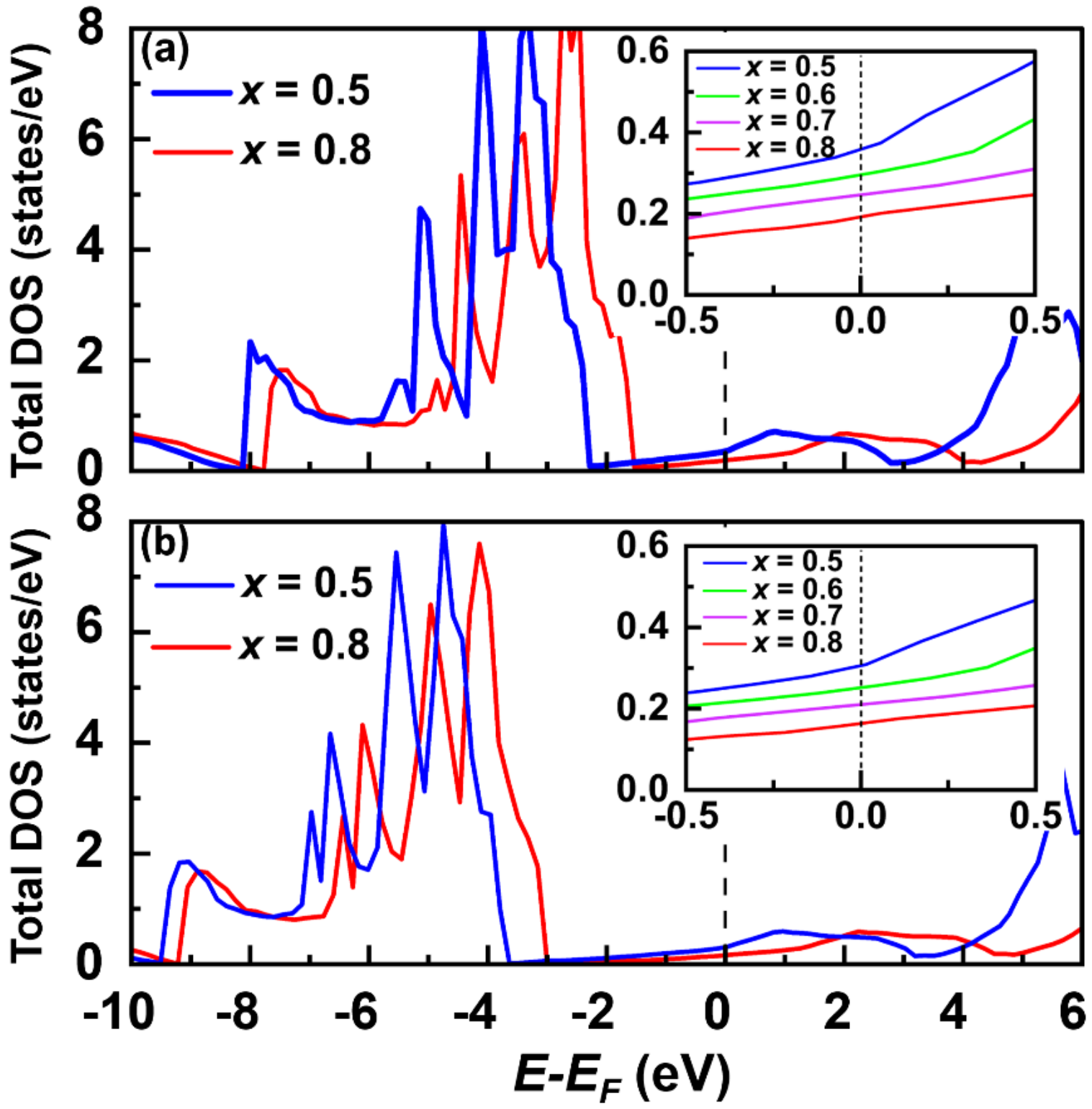

FIG. 6. Total electronic density of states of $Ba_{1-x}K_xSbO_3$ ($x$ = 0.5, 0.8) by using both (a) LDA and (b) HSE06 hybrid functionals. The insets show the electronic DOS near the Fermi energy.

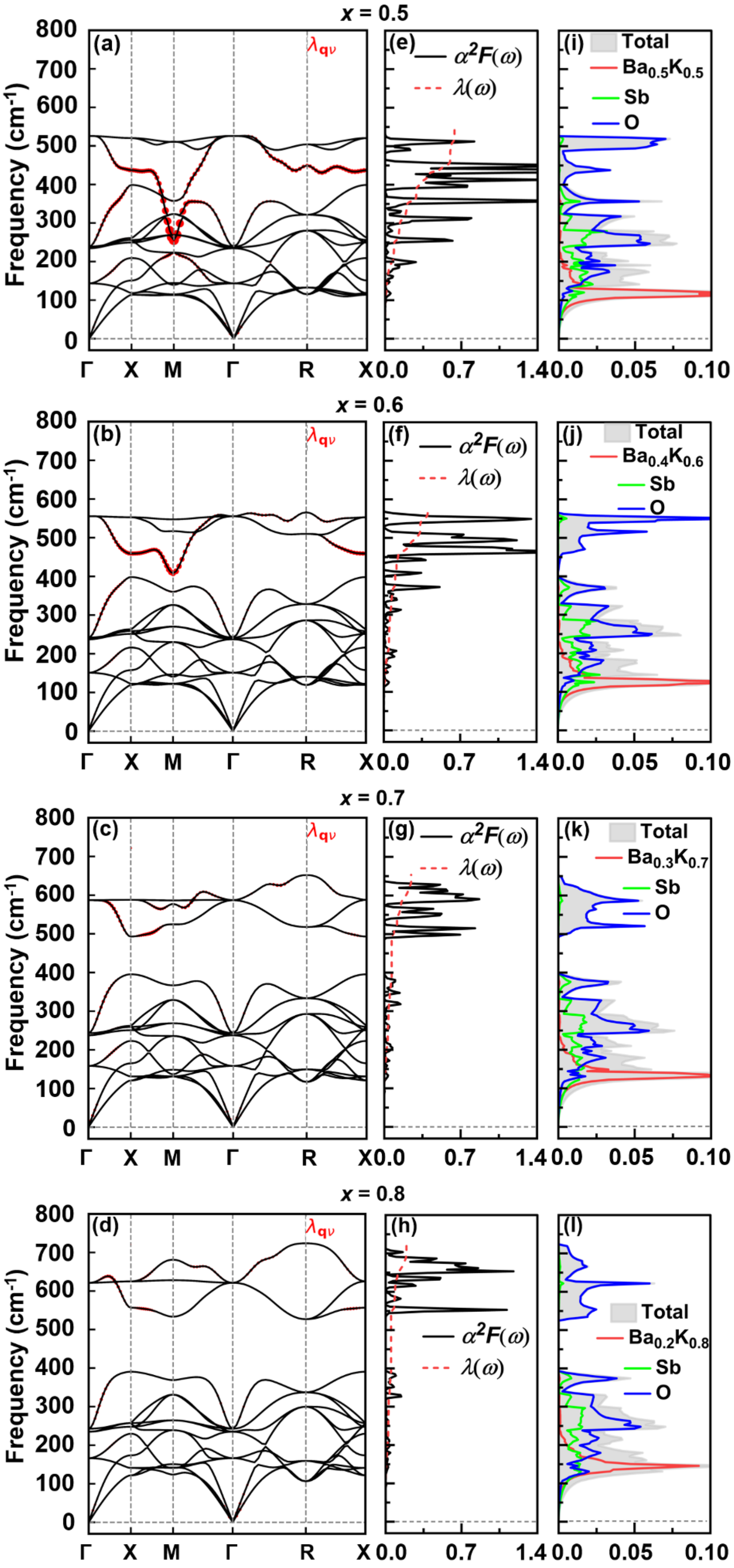

FIG. 7. The DFPT-LDA calculated phonon spectra (left panels (a)-(d), the radius of the red circle is proportional to $\lambda_{qv}$), Eliashberg function $\alpha^2F(\omega)$ (middle panels (e)-(h)) and phonon density of states (right panels (i)-(l)) of $Ba_{1-x}K_xSbO_3$ ($x$ = 0.5, 0.6, 0.7 and 0.8).

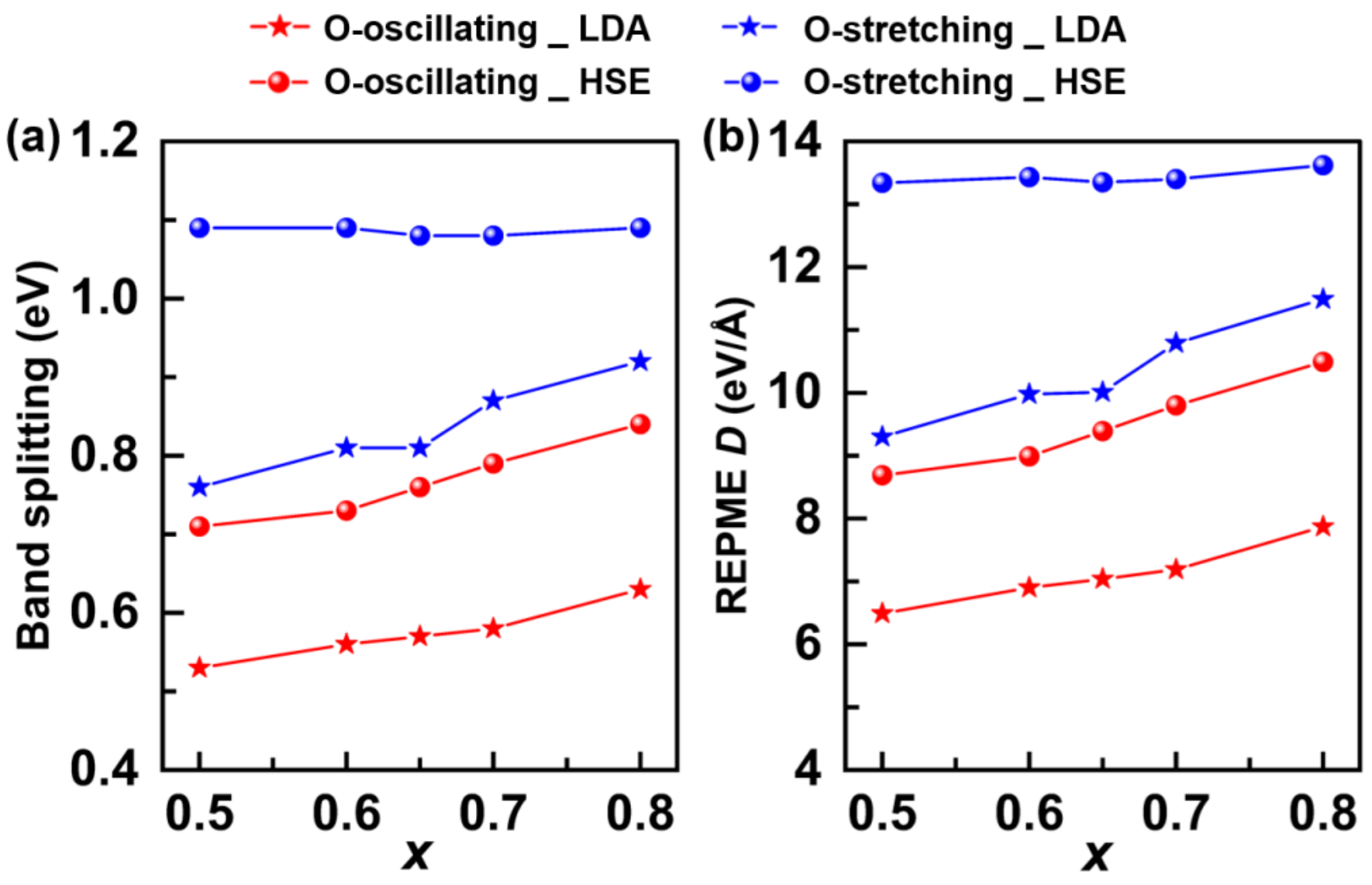

FIG. 8. (a) Band splittings and corresponding REPMEs for the most important vibration modes (oxygen-oscillating mode at the X point and oxygen-stretching mode at the M point) in the LDA and the HSE06 hybrid functionals for $Ba_{1-x}K_xSbO_3$ ($x$ = 0.5, 0.6, 0.65, 0.7, 0.8).

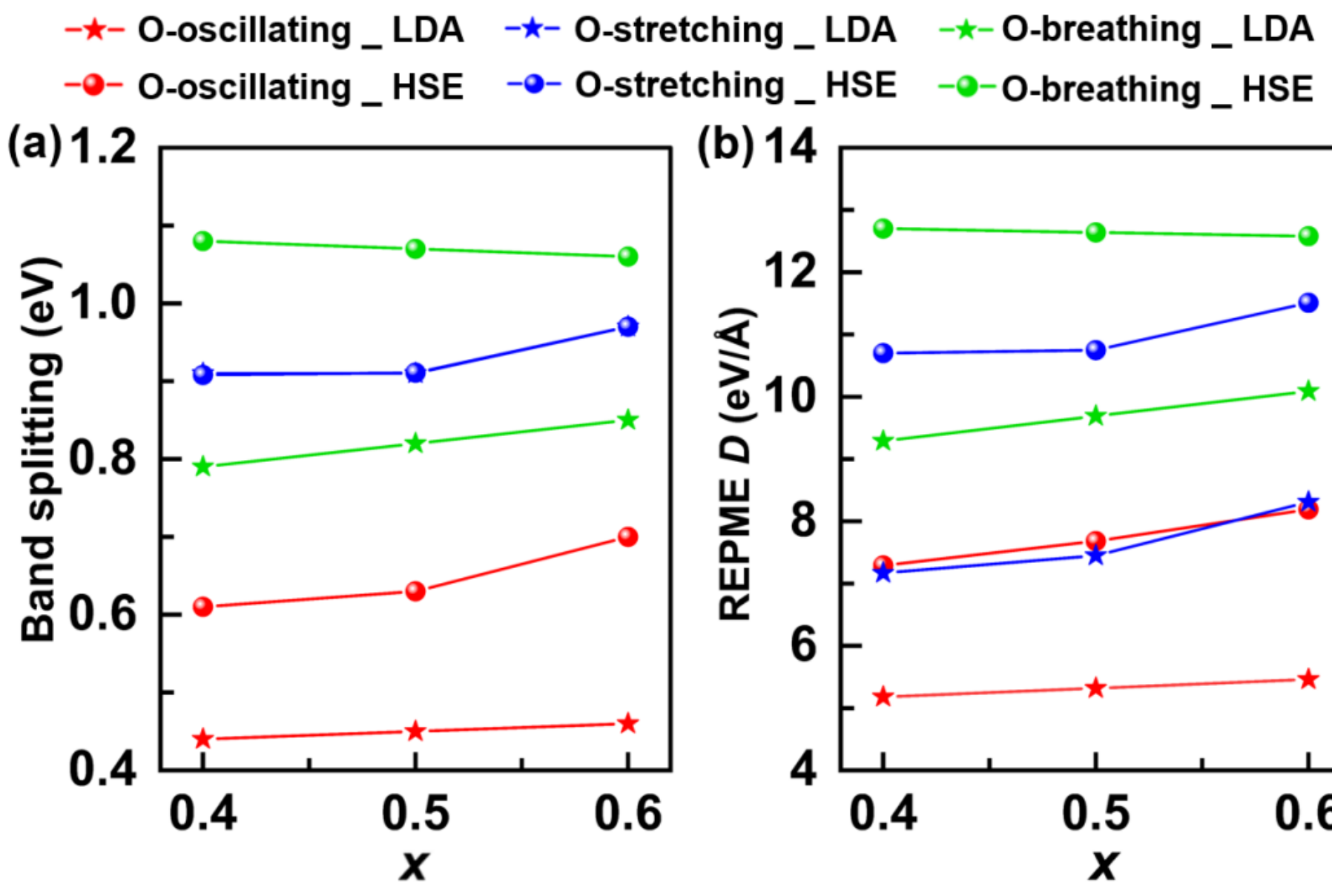

FIG. 9. (a) Band splittings and corresponding REPMEs for the most important vibration modes (oxygen-oscillating mode at the X point, oxygen-stretching mode at the M point and oxygen-breathing mode at the R point) in the LDA and the HSE06 hybrid functionals for $Ba_{1-x}K_xBiO_3$ ($x$ = 0.4, 0.5, 0.6).

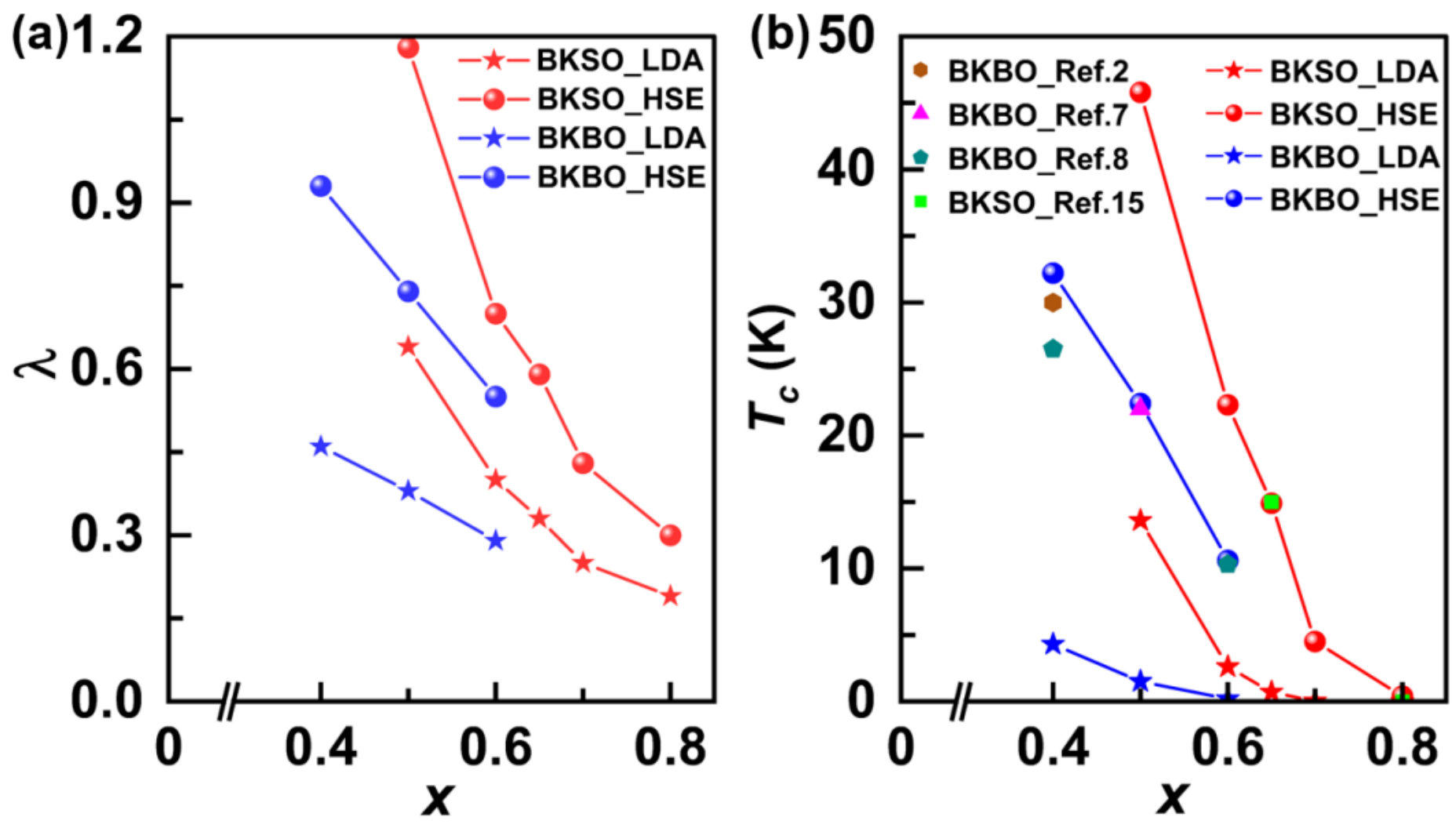

FIG. 10. (a) The total EPC $\lambda$, and critical temperature $T_c$ (K) calculated by the LDA and the HSE06 hybrid functionals as well as the experimental results for $Ba_{1-x}K_xBiO_3$ ($x$ = 0.4, 0.5, 0.6) and $Ba_{1-x}K_xSbO_3$ ($x$ = 0.5, 0.6, 0.65, 0.7, 0.8).

TABLE I. The lattice parameters (Å) of $Ba_{1-x}K_xSbO_3$ ($x$ = 0.5, 0.6, 0.65, 0.7, 0.8) and $Ba_{1-x}K_xBiO_3$ ($x$ = 0.4, 0.5, 0.6) in the cubic perovskite phase.

| $x$ | 0.4 | 0.5 | 0.6 | 0.65 | 0.7 | 0.8 | Ref. |
|---|---|---|---|---|---|---|---|
| $Ba_{1-x}K_xSbO_3$ | 4.101 | 4.076 | 4.050 | 4.037 | 4.023 | 3.997 | VASP |
| | 4.110 | 4.085 | 4.059 | 4.046 | 4.031 | 4.002 | QE |
| | | 4.108 | | 4.067 | | | Ref. [15] |
| $Ba_{1-x}K_xBiO_3$ | 4.251 | 4.231 | 4.213 | | | | VASP |
| | 4.195 | 4.176 | 4.157 | | | | QE |
| | 4.284 | | 4.255 | | | | Ref. [8] |

TABLE II. The bandwidths (eV) along Γ-X direction and Γ-M direction in the LDA and the HSE06 hybrid functionals for $Ba_{1-x}K_xSbO_3$ ($x$ = 0.5, 0.6, 0.65, 0.7, 0.8).

| $x$ | Γ-X | | | Γ-M | | |
|---|---|---|---|---|---|---|
| | LDA | HSE06 | broadening | LDA | HSE06 | broadening |
| 0.5 | 3.76 | 5.24 | 39% | 6.07 | 7.96 | 31% |
| 0.6 | 3.96 | 5.37 | 36% | 6.30 | 8.12 | 29% |
| 0.65 | 4.06 | 5.35 | 32% | 6.42 | 8.12 | 26% |
| 0.7 | 4.16 | 5.32 | 28% | 6.54 | 8.11 | 24% |
| 0.8 | 4.40 | 5.25 | 19% | 6.82 | 8.07 | 18% |

TABLE III. The band splittings (eV), and REPMEs $D$ (eV/Å) for the most important vibration modes in the LDA and the HSE06 hybrid functionals for $Ba_{1-x}K_xSbO_3$ ($x$ = 0.5, 0.6, 0.65, 0.7, 0.8).

| Mode | $x$ | Band splitting | | $D_L$ | $D_H$ | $\frac{\lvert D_H^\nu\rvert^2}{\lvert D_L^\nu\rvert^2}$ |
|---|---|---|---|---|---|---|
| | | LDA | HSE06 | LDA | HSE06 | |
| O breathing at R | 0.5 | 0.87 | 1.12 | 10.65 | 13.71 | 1.66 |
| | 0.5 | 0.53 | 0.71 | 6.49 | 8.69 | 1.79 |
| | 0.6 | 0.56 | 0.73 | 6.90 | 8.99 | 1.70 |
| O stretching at X | 0.65 | 0.57 | 0.76 | 7.04 | 9.39 | 1.78 |
| | 0.7 | 0.58 | 0.79 | 7.19 | 9.80 | 1.86 |
| | 0.8 | 0.63 | 0.84 | 7.87 | 10.49 | 1.78 |
| | 0.5 | 0.76 | 1.09 | 9.30 | 13.34 | 2.06 |
| | 0.6 | 0.81 | 1.09 | 9.98 | 13.43 | 1.81 |
| O stretching at M | 0.65 | 0.81 | 1.08 | 10.01 | 13.35 | 1.78 |
| | 0.7 | 0.87 | 1.08 | 10.79 | 13.40 | 1.54 |
| | 0.8 | 0.92 | 1.09 | 11.49 | 13.62 | 1.41 |

TABLE IV. The enhancement factor of total $\lambda$ by HSE06 hybrid functional, the total EPC $\lambda$, the average phonon frequency $\omega_{\log}$ (K), and the calculated $T_c$ (K) in the LDA and the HSE06 hybrid functionals as well as the experimental $T_c$ (experiment from Ref [15]) for $Ba_{1-x}K_xSbO_3$ (x = 0.5, 0.6, 0.65, 0.7, 0.8).

| $x$ | $\left\langle \frac{\lvert D_H^\nu \rvert^2}{\lvert D_L^\nu \rvert^2} \right\rangle$ | $\lambda$ | | $\omega_{\log}$ | | $T_c$ | | |
|---|---|---|---|---|---|---|---|---|
| | | LDA | HSE06 | LDA | HSE06 | LDA | HSE06 | experiment |
| 0.5 | 1.84 | 0.64 | 1.18 | 502 | 435 | 13.6 | 45.8 | |
| 0.6 | 1.76 | 0.40 | 0.70 | 589 | 535 | 2.6 | 22.3 | |
| 0.65 | 1.78 | 0.33 | 0.59 | 623 | 572 | 0.7 | 14.9 | 15 |
| 0.7 | 1.70 | 0.25 | 0.43 | 644 | 602 | 0.03 | 4.5 | |
| 0.8 | 1.6 | 0.19 | 0.30 | 708 | 677 | 0 | 0.4 | 0 |

TABLE V. The band splittings (eV), and REPMEs $D$ (eV/Å) for the most important vibration modes in the LDA and the HSE06 hybrid functionals for $Ba_{1-x}K_xBiO_3$. ($x$ = 0.4, 0.5, 0.6).

| Mode | $x$ | Band splitting | | $D_L$ | $D_H$ | $\frac{\|D_H^\nu\|^2}{\|D_L^\nu\|^2}$ |
|---|---|---|---|---|---|---|
| | | LDA | HSE06 | LDA | HSE06 | |
| O breathing at R | 0.4 | 0.79 | 1.08 | 9.29 | 12.70 | 1.87 |
| | 0.5 | 0.82 | 1.07 | 9.69 | 12.64 | 1.70 |
| | 0.6 | 0.85 | 1.06 | 10.09 | 12.58 | 1.55 |
| O stretching at X | 0.4 | 0.44 | 0.62 | 5.18 | 7.29 | 1.98 |
| | 0.5 | 0.45 | 0.65 | 5.32 | 7.68 | 2.08 |
| | 0.6 | 0.46 | 0.69 | 5.46 | 8.19 | 2.25 |
| O stretching at M | 0.4 | 0.61 | 0.91 | 7.17 | 10.70 | 2.23 |
| | 0.5 | 0.63 | 0.91 | 7.45 | 10.75 | 2.08 |
| | 0.6 | 0.70 | 0.97 | 8.31 | 11.51 | 1.92 |

TABLE VI. The enhancement factor of total $\lambda$ by HSE06 hybrid functional, the total EPC $\lambda$, the average phonon frequency $\omega_{\log}$ (K), and the calculated $T_c$ (K) in the LDA and the HSE06 hybrid functionals as well as the experimental $T_c$ for $Ba_{1-x}K_xBiO_3$ ($x$ = 0.4, 0.5, 0.6).

| $x$ | $\left\langle\frac{\|D_H^\nu\|^2}{\|D_L^\nu\|^2}\right\rangle$ | $\lambda$ | | $\omega_{\log}$ | | $T_c$ | | |
|---|---|---|---|---|---|---|---|---|
| | | LDA | HSE06 | LDA | HSE06 | LDA | HSE06 | experiment |
| 0.4 | 2.03 | 0.46 | 0.93 | 498 | 433 | 4.3 | 32.2 | 26.5[a], 30[b] |
| 0.5 | 1.95 | 0.38 | 0.74 | 530 | 472 | 1.5 | 22.4 | 22[c] |
| 0.6 | 1.91 | 0.29 | 0.55 | 560 | 511 | 0.2 | 10.6 | 10.3[a] |

[a] Ref. [8], [b] Ref. [2], [c] Ref. [7]

## ACKNOWLEDGMENTS

This work was supported by the National Natural Science Foundation of China (Grant No. 12074041 and 11674030), the Fundamental Research Funds for the Central Universities (Grant No.310421113), the National Key Research and Development Program of China through Contract No. 2016YFA0302300, and the start-up funding of Beijing Normal University. The calculations were carried out with high performance computing cluster of Beijing Normal University in Zhuhai.

*Requests for materials should be addressed to Z.P.Y. at yinzhiping@bnu.edu.cn

# Supplementary Material for "Correlation-enhanced electron-phonon coupling and superconductivity in (Ba,K)$SbO_3$ superconductors"

Zhihong Yuan, Pengyu Zheng, Yiran Peng, Rui Liu, Xiaobo Ma,

Guangwei Wang, Tianye Yu, and Zhiping Yin*

*Department of Physics and Center for Advanced Quantum Studies, Beijing Normal University, Beijing 100875, China.*

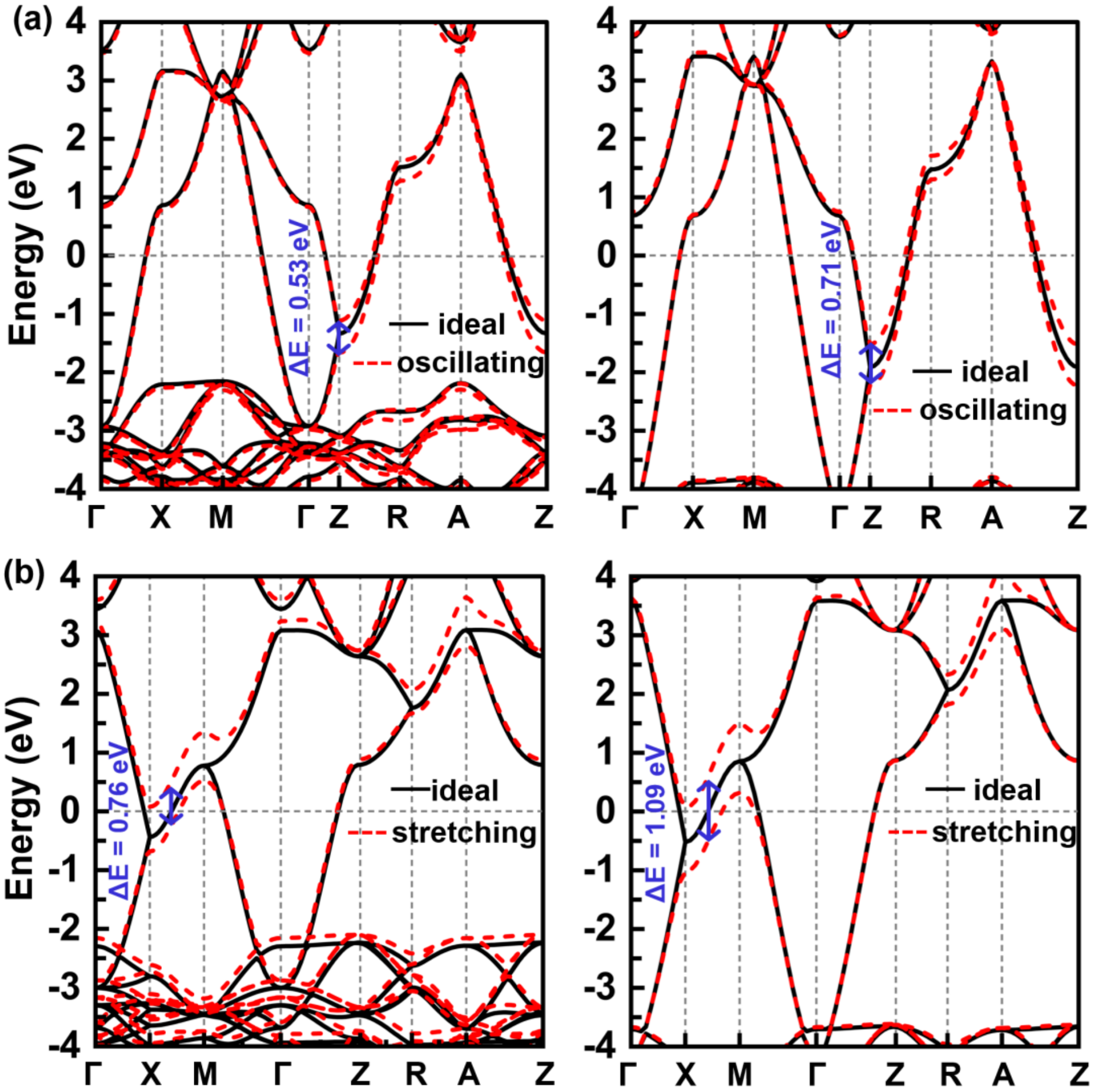

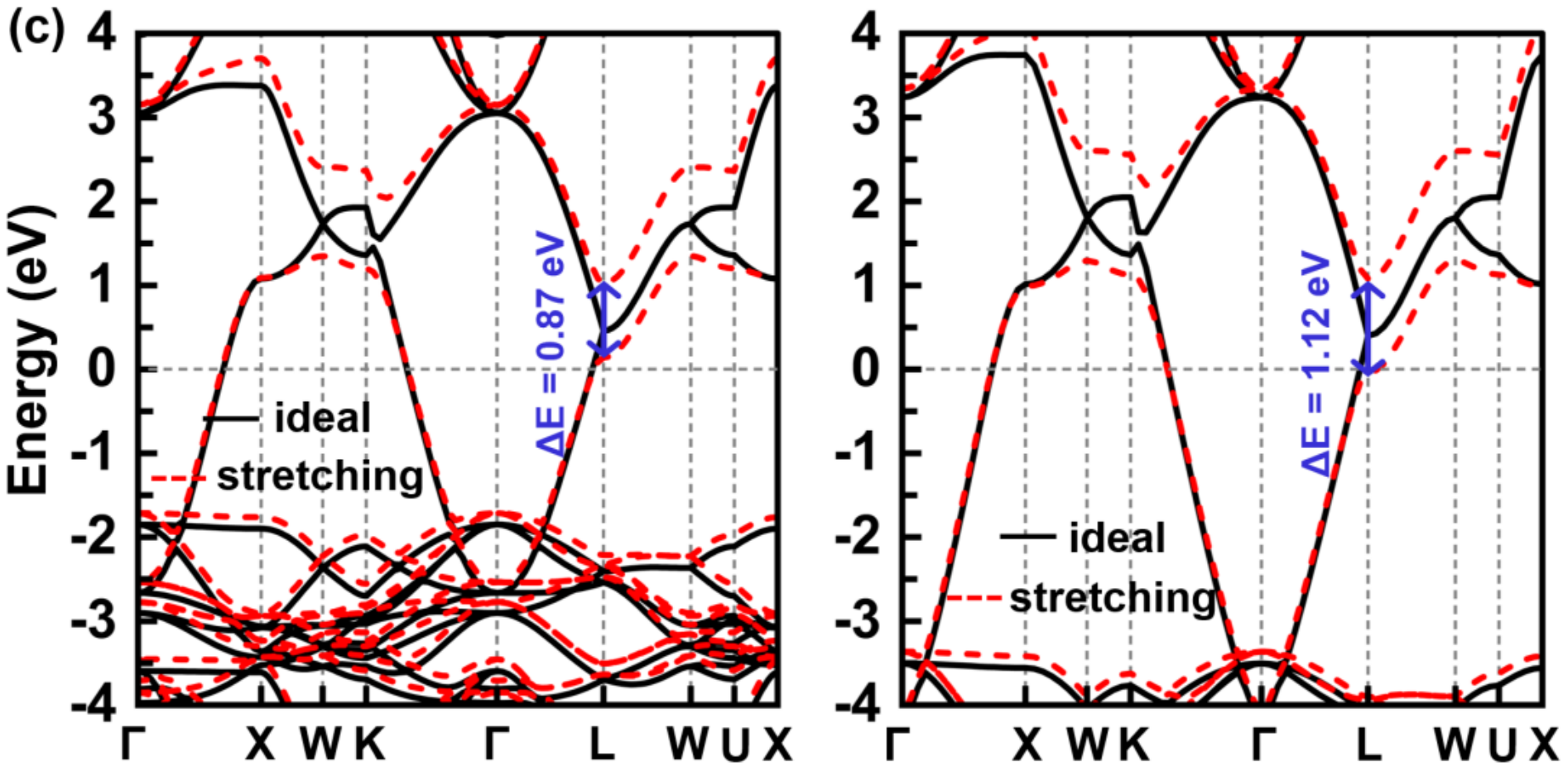

FIG. S1. Illustration of the REPMEs of (a) oxygen-oscillating mode at the X point, (b) oxygen-stretching mode at the M point, and (c) oxygen-breathing mode at the R point in $Ba_{0.5}K_{0.5}SbO_3$. The band structures with and without the oxygen displacement are calculated by DFT using both LDA (left panel) and HSE06 (right panel) hybrid functionals.

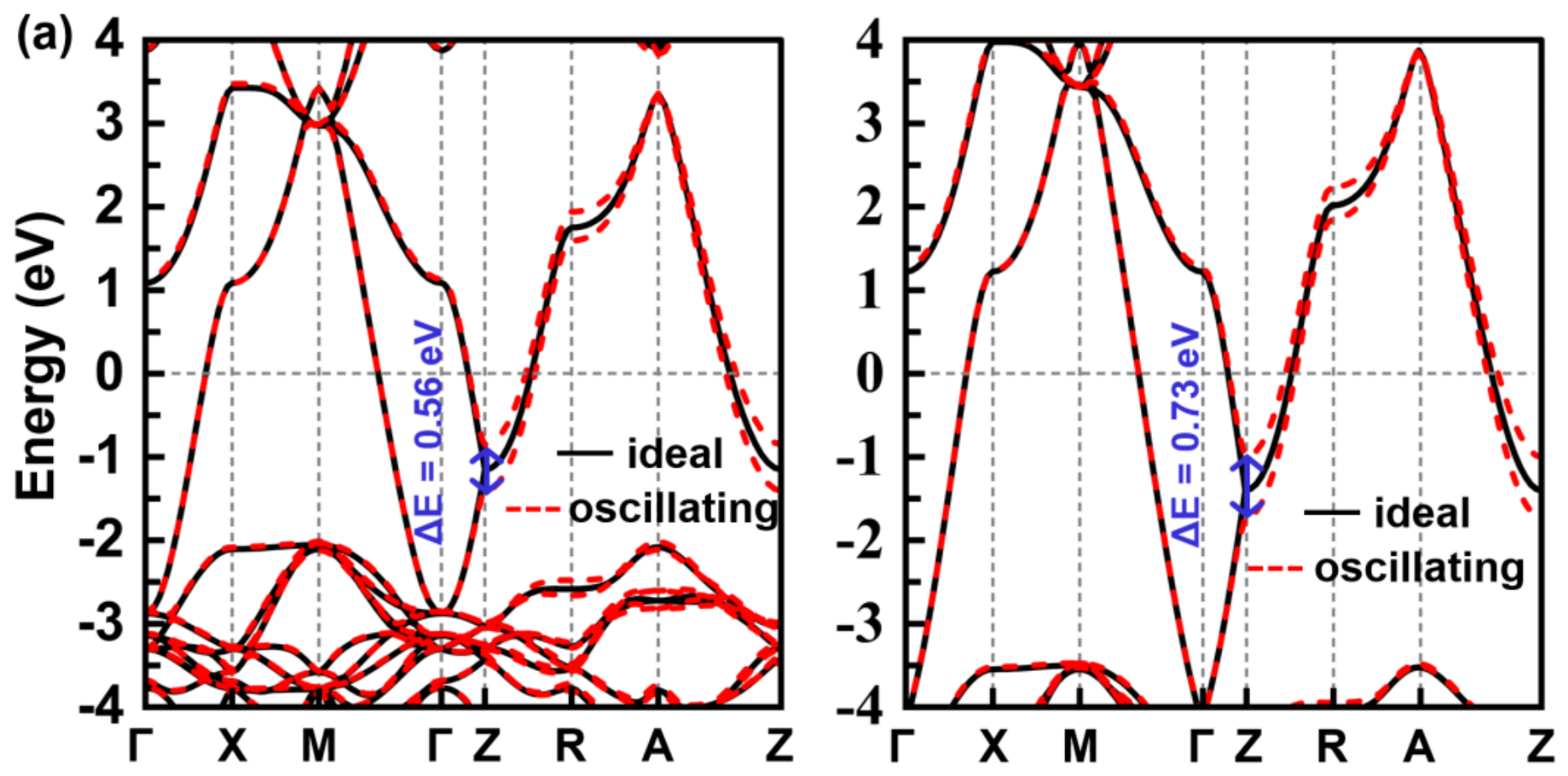

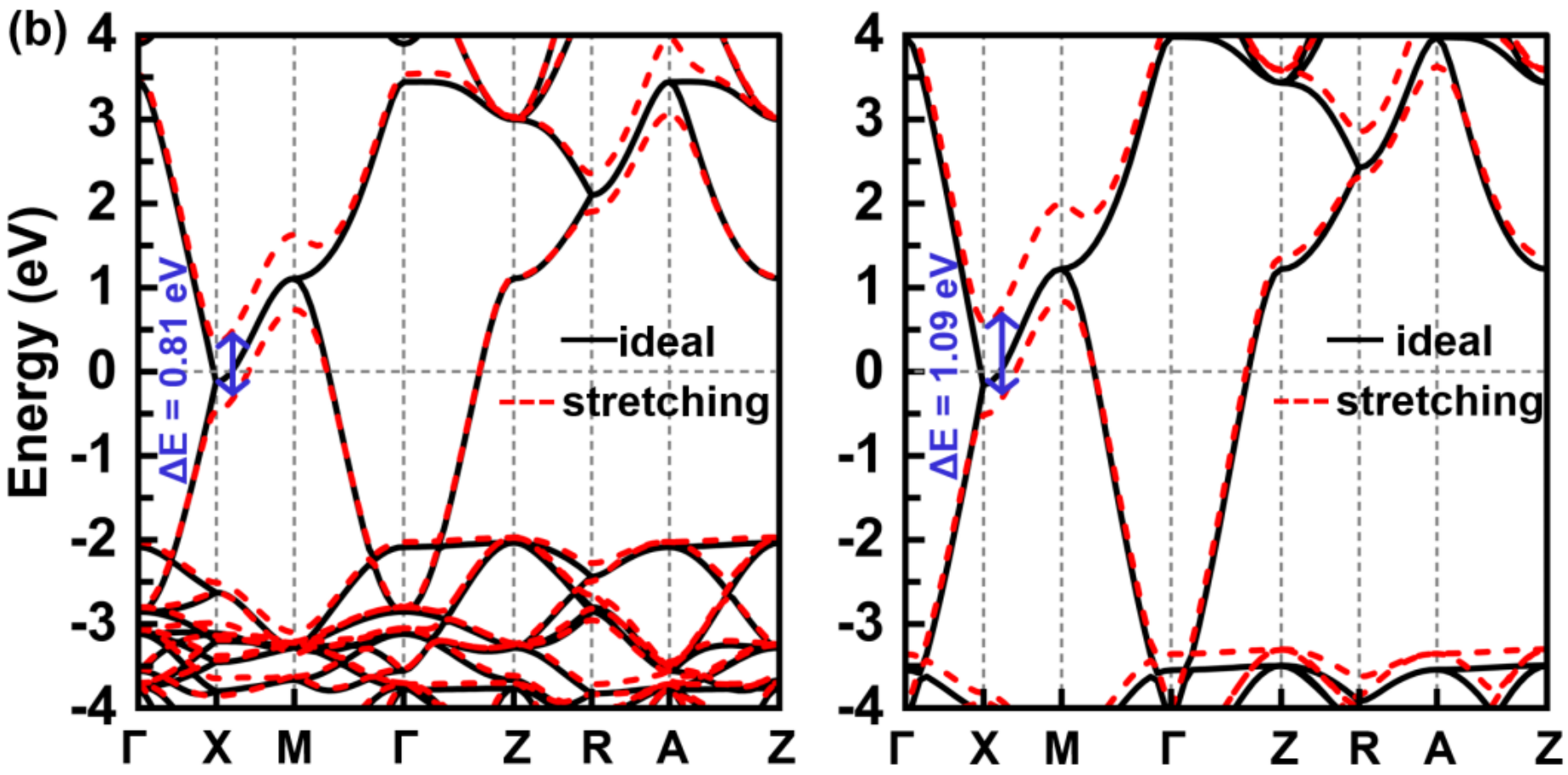

FIG. S2. Illustration of the REPMEs of (a) oxygen-oscillating mode at the X point and (b) oxygen-stretching mode at the M point in $Ba_{0.4}K_{0.6}SbO_3$. The band structures with and without the oxygen displacement are calculated by DFT using both the LDA (left panel) and HSE06 (right panel) hybrid functionals.

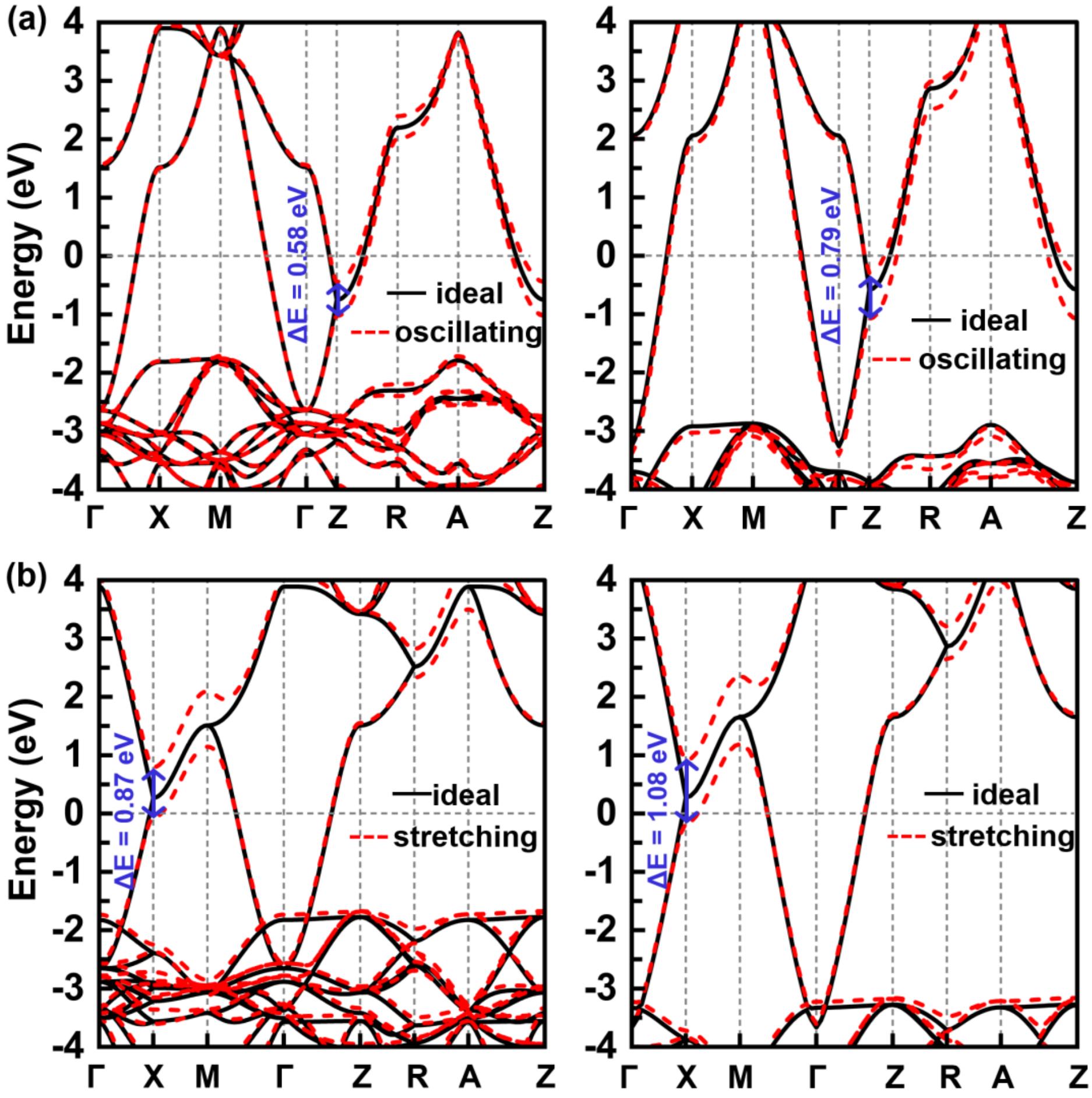

FIG. S3. Illustration of the REPMEs of (a) oxygen-oscillating mode at the X point and (b) oxygen-stretching mode at the M point in $Ba_{0.3}K_{0.7}SbO_3$. The band structures with and without the oxygen displacement are calculated by DFT using both the LDA (left panel) and HSE06 (right panel) hybrid functionals.

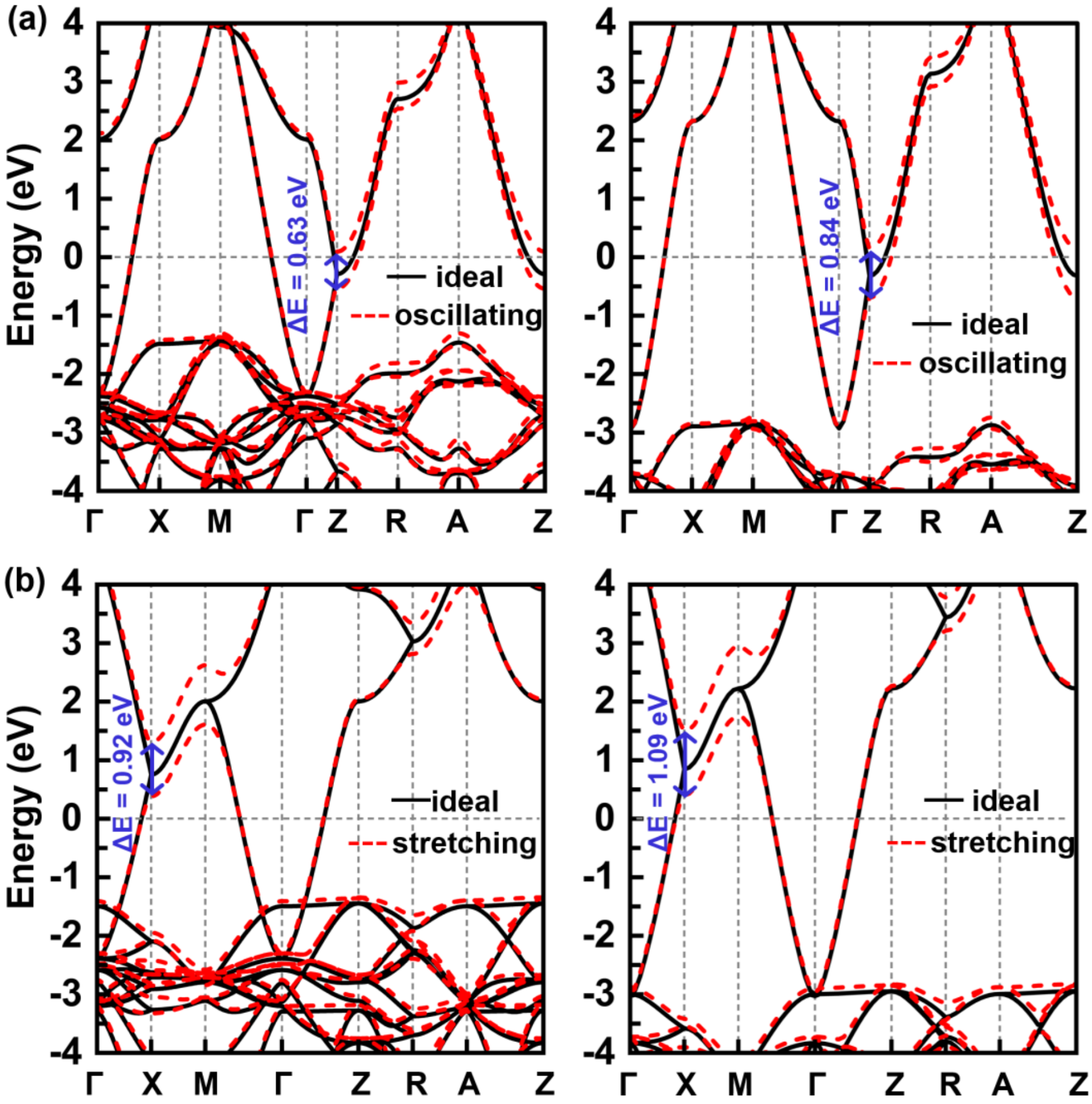

FIG. S4. Illustration of the REPMEs of (a) oxygen-oscillating mode at the X point and (b) oxygen-stretching mode at the M point in $Ba_{0.2}K_{0.8}SbO_3$. The band structures with and without the oxygen displacement are calculated by DFT using both the LDA (left panel) and HSE06 (right panel) hybrid functionals.